\documentclass[graybox]{SNmult}

\usepackage{type1cm}         
\usepackage{makeidx}         
\usepackage{graphicx}        
\usepackage{multicol}        
\usepackage[bottom]{footmisc}

\usepackage{newtxtext}
\usepackage[varvw]{newtxmath}

\usepackage{booktabs}        
\usepackage{textcomp}        
\usepackage{float}

\makeindex

\newcommand{\miniKapitel}[1]{\noindent\textit{#1:}\ }

\begin{document}
\title*{POI Recommendation with LLM-Augmented Multi-Graph Learning and Contrastive Alignment}
\titlerunning{POI Recommendation with LLM-Augmented Multi-Graph Learning and CA}
\author{Burak Tamer, Wolfram H\"opken, Zehui Wang and Thomas Bayer}
\authorrunning{B. Tamer et al.}
\institute{Burak Tamer, Wolfram H\"opken, Zehui Wang and Thomas Bayer \at Institute for Digital Transformation - University of Applied Sciences Ravensburg-Weingarten, Germany,\\ \email{burak.tamer@rwu.de}, \email{wolfram.hoepken@rwu.de}, \email{zehui.wang@rwu.de}, \email{thomas.bayer@rwu.de}}
\maketitle

\abstract*{Point-of-interest (POI) recommendation models based on graph neural networks achieve strong performance by propagating collaborative signals over user-item interactions, yet they struggle with the cold-start problem, where items with few or no interactions are not represented. In this paper, we propose LLM-augmented Multi-Graph Contrastive Learning (LLM-MGCL), a multi-graph neural network that uses semantic and spatial information about items to extend the LightGCN backbone with two auxiliary item-item graphs: a semantic graph constructed from sentence embeddings of LLM-generated photo summaries and keywords, and a geographic graph derived from Haversine distances between business locations. Item embeddings are propagated over all three graphs in parallel, fused additively, and aligned across views through a bidirectional InfoNCE contrastive objective that connects behavioral, semantic, and spatial representations of the same items. Experiments on the Yelp Multimodal Recommendation Dataset show that LLM-MGCL outperforms classical collaborative filtering, matrix factorization, and non-contrastive graph neural network baselines. It improves Recall@20 by 48.5\% and NDCG@20 by 59.0\% over LightGCN, while trailing the strongest contrastive baseline, Self-supervised Graph Learning (SGL), slightly in aggregate. A popularity-stratified evaluation shows that LLM-MGCL achieves the highest recall on cold and warm items, whereas SGL remains better on popular items. An ablation study reveals that the cross-view contrastive alignment (CA) is the primary driver of the aggregate gains. Our results suggest that externally grounded, LLM-derived item knowledge can partially compensate for missing collaborative signal and mitigate the item cold-start problem in POI recommendation.
}

\abstract{Point-of-interest (POI) recommendation models based on graph neural networks achieve strong performance by propagating collaborative signals over user-item interactions, yet they struggle with the cold-start problem, where items with few or no interactions are not represented. In this paper, we propose LLM-augmented Multi-Graph Contrastive Learning (LLM-MGCL), a multi-graph neural network that uses semantic and spatial information about items to extend the LightGCN backbone with two auxiliary item-item graphs: a semantic graph constructed from sentence embeddings of LLM-generated photo summaries and keywords, and a geographic graph derived from Haversine distances between business locations. Item embeddings are propagated over all three graphs in parallel, fused additively, and aligned across views through a bidirectional InfoNCE contrastive objective that connects behavioral, semantic, and spatial representations of the same items. Experiments on the Yelp Multimodal Recommendation Dataset show that LLM-MGCL outperforms classical collaborative filtering, matrix factorization, and non-contrastive graph neural network baselines. It improves Recall@20 by 48.5\% and NDCG@20 by 59.0\% over LightGCN, while trailing the strongest contrastive baseline, Self-supervised Graph Learning (SGL), slightly in aggregate. A popularity-stratified evaluation shows that LLM-MGCL achieves the highest recall on cold and warm items, whereas SGL remains better on popular items. An ablation study reveals that the cross-view contrastive alignment (CA) is the primary driver of the aggregate gains. Our results suggest that externally grounded, LLM-derived item knowledge can partially compensate for missing collaborative signal and mitigate the item cold-start problem in POI recommendation.
}

\section{Introduction}\label{sec1}
Point-of-interest (POI) recommendation has become a central component of location-based services, helping users navigate and discover businesses that match their preference and location \cite{zhang2025poisurvey}. Graph neural networks (GNNs), and Light Graph Convolution Network (LightGCN) \cite{he2020lightgcn} in particular, have emerged as a strong backbone for this task by propagating collaborative signals across user and item interactions, becoming a powerful paradigm enabling state-of-the-art performance \cite{WugraphneuralnetworksSurvey, he2020lightgcn, Wang2019ngcf}. However, purely collaborative approaches rely exclusively on observed interaction patterns and therefore struggle with the cold-start problem \cite{schein2002coldstartproblem}: businesses with few or no interactions by users receive little to no useful signal during training and are therefore not recommended. At the same time, semantic and spatial information about businesses, such as what they look like, what services they offer, and where they are located, is available but rarely exploited by these models. Since this information does not depend on the user or the interaction, but rather on the businesses, it could directly compensate for the missing collaborative signal that limits recommendation in the cold-start problem.

In this work, we propose to leverage LLM-generated textual descriptions of businesses from multimodal data (images and metadata), as well as geographic proximity to enhance POI recommendation. The central research question is whether incorporating LLM-generated textual information such as summaries, keywords and spatial relationships between businesses into a multi-graph neural network with contrastive alignment can improve recommendation performance compared to state-of-the-art approaches that rely solely on user-item interactions. The main contributions of this work are as follows:
\begin{itemize}
    \item{Construction of a semantic item-item graph based on textual embeddings derived from LLM-generated descriptions and keywords, capturing semantic similarity between businesses in the Yelp Multimodal Recommendation Dataset \cite{wang2025llmaugmented}.}
    \item{Construction of a geographic item-item graph based on the Haversine distance between business locations, capturing spatial proximity that is particularly relevant for POI recommendation.}
    \item{Design of a multi-graph model (LLM-MGCL) that integrates user-item interactions with both item-item graphs, and its evaluation against collaborative filtering (CF), matrix factorization (MF) and interaction-only graph baselines.}
 \end{itemize}
The remainder of this paper is structured as follows. Section~\ref{sec2} reviews related work in recommendation systems, graph neural networks and the use of  textual information for recommendation. In Sect.~\ref{sec3} we describe the dataset used, the proposed methodology and model architecture, including graph construction and the GNN training procedure. Lastly, Sect.~\ref{sec4} presents the experimental results compared to other state-of-the-art models and Sect.~\ref{sec5} concludes the paper and discusses potential future work.

\section{Related Work}\label{sec2}
Recommender systems have evolved from modeling user-item interactions to increasingly incorporating richer sources of information, such as auxiliary graph structure, self-supervised training signals, and external content beyond the interaction history itself \cite{WugraphneuralnetworksSurvey}. Each development addresses a specific limitation of purely interaction-based collaborative filtering, and together they motivate the design choices underlying LLM-MGCL.

\miniKapitel{GNN-based Recommendation} Graph neural networks (GNNs) have become a dominant paradigm for collaborative filtering by modeling user-item interactions in a bipartite graph structure \cite{WugraphneuralnetworksSurvey}. NGCF \cite{Wang2019ngcf} introduced standard graph convolution while LightGCN \cite{he2020lightgcn} showed that a simplified propagation scheme of only neighborhood aggregation achieves better performance. Related work such as GC-MC \cite{Berg2017gcmc} and PinSage \cite{Ying2018pinsage} similarly demonstrated the effectiveness of graph-based propagation for recommendation systems. These graph-based methods achieve strong  performance, but inherit the same dependence on interaction data discussed in Sect.~\ref{sec1}.

\miniKapitel{Contrastive Learning} To improve the robustness of learned representations, self-supervised contrastive learning has recently been introduced into graph-based recommendation \cite{WugraphneuralnetworksSurvey}. SGL \cite{wu2021SGL} provides an additional training signal beyond the ranking objective by generating two augmented views of the interaction graph through random edge or node dropout and trains the model to align representations of the same node across different views. Recent work such as SimGCL \cite{Yu2022simgcl} and NCL \cite{Lin2022ncl} refined this idea by replacing structural augmentation with simpler embedding level perturbations. A common characteristic of these methods is that the contrasted views are perturbations of the same interaction graph, meaning the additional signal is derived entirely from the same source of information the model already used \cite{wu2021SGL}.

\miniKapitel{Content and Geographic Signals}
A separate line of work incorporates external information, such as text, images, or knowledge graphs, to complement collaborative signals. KGAT \cite{Wang2018kgat}, for example, integrates a knowledge graph of item attributes into the propagation process, while multimodal recommenders combine visual and textual item features into the propagation process \cite{Zhou2023multimodalrec}. With the emergence of large language models (LLMs), recent work has begun incorporating LLM-generated content as a richer source of item semantics, reflecting a trend towards using LLMs for recommender systems \cite{Wu2025llmrecsys}. Geographic proximity has similarly been used as an auxiliary signal in POI recommendation, reflecting the intuition that users are more likely to interact with nearby businesses \cite{zhang2025poisurvey}.

This work builds on the LightGCN backbone and extends it with LLM-generated semantic content and geographic proximity, aligned through a contrastive learning objective. Unlike SGL and its successors, LLM-MGCL contrasts collaborative views against semantic and geographic representations, rather than randomly perturbed views.

\section{Methodology}\label{sec3}
\begin{figure}[h]
\includegraphics[width=\textwidth]{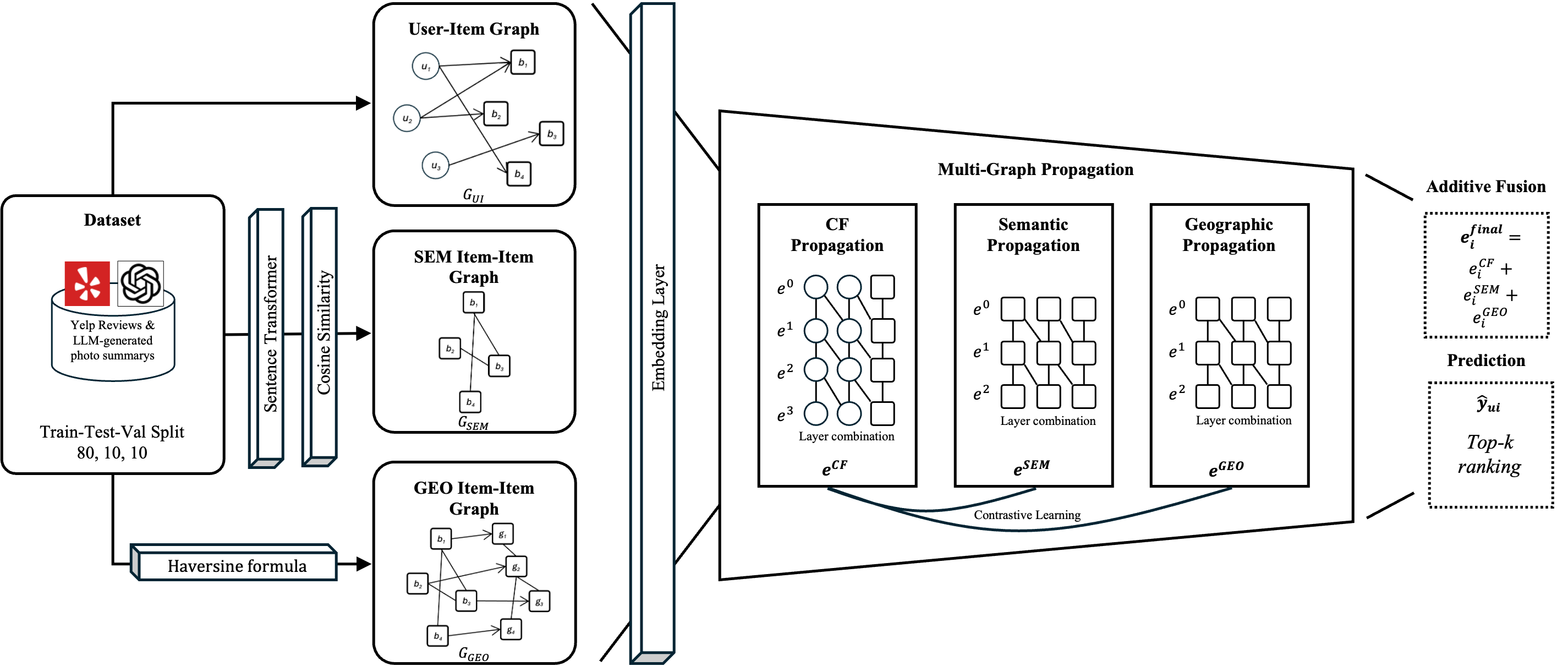}
\Description{Overview diagram of the LLM-MGCL methodology showing the construction of the three input graphs, the parallel propagation paths and the fusion and contrastive alignment of the resulting item representations.}
\caption{Overview of the LLM-MGCL methodology. Source: Authors' own.}
\label{fig:methodik}
\end{figure}

The presented approach combines collaborative filtering signals with LLM-derived semantic and geographic information through a multi-graph architecture. LLM-MGCL extends the LightGCN backbone with two parallel propagation paths over item-item  graphs constructed from LLM-generated semantic similarity and geographic proximity and aligns the resulting item representations through contrastive learning objectives.
As illustrated in Fig.~\ref{fig:methodik}, LLM-MGCL uses three graphs as input, a user-item graph $G_{UI}$ encoding collaborative behavior between users and businesses, an item-item similarity graph $G_{SEM}$ derived from LLM-generated photo summaries and an item-item geographic graph $G_{GEO}$ derived from Haversine distances between business locations. All three graphs share the same item node set, allowing semantic, behavioral and geographic signals to influence the same underlying item representation.

\noindent The following sections describe the dataset (Sect.~\ref{sec3.1}), the construction of each graph (Sect.~\ref{sec3.2}), the model architecture (Sect.~\ref{sec3.3}) and the training procedure (Sect.~\ref{sec3.4}).

\subsection{Dataset and Split}\label{sec3.1}
Experiments are conducted on the Yelp Multimodal Recommendation Dataset, which is derived from the online business review platform Yelp and widely adopted in research on collaborative filtering and POI recommendation \cite{WugraphneuralnetworksSurvey}. Each interaction consists of a user ID, a business ID and a star rating; in addition, the dataset provides multimodal information about businesses such as images and metadata. 

We use the processed and extended version provided by Wang et al. \cite{wang2025llmaugmented} on Hugging Face\footnote{\url{https://huggingface.co/datasets/wzehui/Yelp-Multimodal-Recommendation}}, in which the LLM ChatGPT transform this multimodal information into natural language summaries and keywords, providing a high-level semantic representation of each business \cite{wang2025llmaugmented}.

\miniKapitel{Train-Test-Validation Split} We apply a per-user random split, retaining only users with at least three interactions so that every user can contribute to all partitions. Per user, interactions are shuffled with a fixed random seed and divided into training ($80\%$), validation ($10\%$) and test ($10\%$) sets.

\subsection{Graph construction}\label{sec3.2}
LLM-MGCL operates on three complementary graphs that capture different types of relationships between users and businesses. The user-item graph $G_{UI}$ encodes observed collaborative behavior, while the two item-item graphs $G_{SEM}$ and $G_{GEO}$ provide auxiliary structural information, with $G_{SEM}$ derived from LLM-generated semantic content and $G_{GEO}$ from geographic locations.

\subsubsection{User-Item Graph}\label{sec3.2.1}
The user-item graph $G_{UI} = (\mathcal{U} \cup \mathcal{I}, \mathcal{E}_{UI})$ is a bipartite graph where $\mathcal{U}$ denotes the set of users, $\mathcal{I}$ the set of businesses and $\mathcal{E}_{UI}$ the set of observed user-business interactions. Each interaction corresponds to a review submitted by a user $u \in \mathcal{U}$ for a business $i \in \mathcal{I}$, accompanied by a star rating $r_{ui} \in \{1, 2, 3, 4, 5\}$. To incorporate star rating information into the graph structure, each edge $(u, i) \in \mathcal{E}_{UI}$ is assigned a weight $w_{ui}$ derived from the user's star rating.
\begin{equation}
w_{ui} = \max\left(\frac{r_{ui} - 1}{4},\ 0.1\right)
\label{eq:starweighting}
\end{equation}
This transformation (Eq.~\ref{eq:starweighting})  maps 5-star ratings to a maximum weight of 1.0 and 1-star ratings to a minimum weight of 0.1, which maps $r_{ui} \in \{1, 2, 3, 4, 5\}$ to $w_{ui} \in \{0{.}1, 0{.}25, 0{.}5, 0{.}75, 1{.}0\}$. The lower clamp of 0.1 ensures that low-rated interactions still contribute to message passing, preventing complete elimination of low-rated reviews while emphasizing strong positive ratings. The resulting graph is converted into a symmetric adjacency matrix $\mathbf{A}_{UI} \in \mathbb{R}^{(|\mathcal{U}|+|\mathcal{I}|) \times (|\mathcal{U}|+|\mathcal{I}|)}$ which contains the weighted interactions. The matrix is then normalized using the standard LightGCN formulation \cite[Eq.~8]{he2020lightgcn}. This symmetric normalization stabilizes message passing by counteracting the effect of high-degree nodes during propagation \cite[p.~4-5]{he2020lightgcn}. After preprocessing, $G_{UI}$ contains $|\mathcal{U}| = 22{,}207$ users, $|\mathcal{I}| = 16{,}702$ businesses and $1{,}247{,}788$ bidirectional weighted edges, corresponding to $623{,}894$ unique user-business interactions in the training split.

\subsubsection{SEM-Item-Item Graph}\label{sec3.2.2}

The SEM-item-item graph $G_{SEM} = (\mathcal{I}, \mathcal{E}_{SEM})$ encodes semantic similarity between businesses based on LLM-generated textual content. For each business, the LLM-generated \textit{photo summary} and \textit{keywords} \cite{wang2025llmaugmented} are concatenated into a single textual representation $t_i$ and encoded with the pretrained Sentence-Transformer model \textit{all-MiniLM-L6-v2} into an L2-normalized embedding $\mathbf{s}_i \in \mathbb{R}^{384}$, so that the cosine similarity $\text{sim}(i,j)$ between two businesses reduces to the inner product $\mathbf{s}_i \cdot \mathbf{s}_j$:
\begin{equation}
\text{sim}(i, j) = \frac{\mathbf{s}_i \cdot \mathbf{s}_j}{\|\mathbf{s}_i\| \cdot \|\mathbf{s}_j\|}
\label{eq:cosinesimilarity}
\end{equation}
To construct a sparse and informative graph, top-k filtering is applied: for each business only the $k = 10$ most similar businesses are retained as neighbors. This bounds the graph density at $\mathcal{O}(k \cdot |\mathcal{I}|)$ edges and prevents items with broadly applicable descriptions from dominating the neighborhood structure. The fixed neighborhood size also guarantees that every item, including those with few or no observed user-item interactions, receives a consistent semantic signal during propagation, directly addressing the item cold-start problem inherent to collaborative filtering \cite{schein2002coldstartproblem}. Edges are added bidirectionally with the similarity score as edge weight, and the resulting adjacency matrix is normalized just like for $G_{UI}$. The final graph contains $265{,}780$ bidirectional weighted edges with a mean raw similarity of $0{.}87$, indicating that retained neighbors are semantically close to their target items.

\subsubsection{GEO-Item-Item Graph}\label{sec3.2.3}
The geographic-item-item graph $G_{GEO} = (\mathcal{I}, \mathcal{E}_{GEO})$ encodes spatial proximity between businesses based on their physical locations. While $G_{SEM}$ captures semantic similarity, $G_{GEO}$ captures locational relationships, which are relevant for POI recommendations, where users typically interact with businesses in geographic proximity.

\miniKapitel{Coordinate Extraction} Each business $i \in \mathcal{I}$ is associated with a geographic location represented by latitude $\phi_i$ and longitude $\lambda_i$, both extracted from the Yelp business metadata. 

\miniKapitel{Distance Computation} The geographic distance between two businesses $i$ and $j$ is computed using the Haversine great-circle formula:
\begin{equation}
d(i, j) = 2R \cdot \arcsin\left(\sqrt{\sin^2\left(\frac{\Delta\phi}{2}\right) + \cos(\phi_i)\cos(\phi_j)\sin^2\left(\frac{\Delta\lambda}{2}\right)}\right)
\label{eq:haversineformula}
\end{equation}
The result $d(i,j)$ represents the great-circle distance in kilometers between the two compared businesses.

\miniKapitel{Edge filtering} Analogous to $G_{SEM}$, only pairs within a distance threshold of $d_{max} = 50$ km are considered, and for each business the $k_{geo} = 10$ nearest neighbors are retained, ensuring a balanced neighborhood structure regardless of regional business density.

\miniKapitel{Edge Weighting} Unlike the SEM-item-item graph, where higher cosine similarity directly corresponds to stronger edges, the geographic edge weights are defined by the calculated distances of the Haversine formula. Closer businesses should have stronger connections; therefore, the edge weights are defined as:
\begin{equation}
w_{ij}^{GEO} = \frac{1}{1 + d(i, j)}
\label{eq:GEOgraphweights}
\end{equation}
This formulation ensures that weights decrease smoothly as distance grows and stay bounded between $0$ and $1$, matching the scale of the cosine-based weights in $G_{SEM}$ to keep the signal stable during message passing.

\miniKapitel{Graph Assembly} To form an undirected graph, edges are added bidirectionally. Like every graph, the matrix is symmetrically normalized following the LightGCN scheme. The final GEO-item-item graph contains $334{,}040$ bidirectional weighted edges. Before normalization, the raw edge weights have a mean of $0{.}74$,  reflecting the high spatial density of businesses in the same area, which is particularly well suited for POI recommendations, since users typically choose between alternatives in close proximity.

\subsection{LLM-MGCL Architecture}\label{sec3.3}
LLM-MGCL builds on LightGCN, which learns user and item representations through simple neighborhood aggregation without learned weight matrices or nonlinear activations \cite{he2020lightgcn}. Instead of propagating only over the user-item graph, LLM-MGCL runs three propagation paths in parallel and combines the output into a unified item representation. 

\miniKapitel{Embedding Layer}\label{sec3.3.1}
Before any graph propagation takes place, each user and item is assigned a trainable embedding vector of the dimension $64$. These vectors are the representation of the users and items in a continuous space where similar entities end up close to each other in training. Each user and item embedding matrix is randomly initialized from a zero-mean normal distribution with a standard deviation  of $\sigma = 0.01$ and is updated end-to-end during the training. The item embedding matrix is hereby shared across all propagation paths. This means that when the model updates item representations based on semantic similarity, those changes also affect the same embeddings used in the other paths.

\miniKapitel{Multi-Graph Propagation}\label{sec3.3.2}
The heart of LLM-MGCL is a propagation mechanism that spreads information through each graph.  The mechanism is straightforward: if two items are connected in a graph --- because users interact with both, because they are semantically similar or because they are geographically close, they should influence each other's representation. While LightGCN applies this mechanism exclusively to collaborative filtering propagation, LLM-MGCL spreads information through multiple graphs simultaneously. Expanding this mechanism by stacking multiple propagation layers, the model can aggregate information not just from direct neighbors but also from second- and third-order connections. All three paths use the same propagation formula, following the LightGCN neighborhood aggregation scheme \cite{he2020lightgcn}. At each layer, a node's new representation is a weighted sum of its neighbors' current representations, where the weights are given by the pre-computed normalized adjacency matrix. After all layers are computed, the representations from each layer, including the initial embedding, are averaged to produce the final output. This layer-wise mean pooling ensures that not only local (one-hop) but also global (multi-hop) neighborhood information contribute to the final representation \cite[p.~4]{he2020lightgcn}. The collaborative path propagates the users and item embeddings over $G_{UI}$ for the recommended three layers \cite{he2020lightgcn}. Since $G_{SEM}$ and $G_{GEO}$ contain no user nodes, the semantic and geographic paths propagate only the item embeddings, for two layers each. A shallower depth for these two paths is used because each item is connected only to its k-nearest neighbors. A deeper propagation would mix in loosely related items and dilute rather than sharpen the signal. This results in three separate item representations: a collaborative view $\mathbf{e}_i^{CF}$, a semantic view $\mathbf{e}_i^{SEM}$ and a geographic view $\mathbf{e}_i^{GEO}$.
\\

\miniKapitel{Fusion and Prediction}\label{sec3.3.3}
To combine the three views into a single representation for ranking, LLM-MGCL uses additive fusion, which lets all views contribute equally and keeps the model free of additional learned parameters:
\begin{equation}
\mathbf{e}_i^{\text{final}} = \mathbf{e}_i^{CF} + \mathbf{e}_i^{SEM} + \mathbf{e}_i^{GEO}
\label{eq:additiveFusion}
\end{equation}
The interaction score between a user $u$ and an item $i$ is the dot product $\hat{y}_{ui} = \mathbf{e}_u^{CF} \cdot \mathbf{e}_i^{\text{final}}$; during inference, all candidate items are scored and ranked to produce the top-k recommendation list. A higher score means the model rates the item as more relevant.

\miniKapitel{Contrastive Learning Objective}\label{sec3.3.4}
Without an additional training signal, the three propagation paths are only combined in the fusion step and are not forced to produce consistent representations of the same business. LLM-MGCL addresses this with a contrastive learning objective that explicitly pulls together representations of the same item across different views while pushing apart representations of different items.

In contrast to SGL \cite{wu2021SGL}, which contrasts two randomly perturbed views of
the same interaction graph, LLM-MGCL contrasts fundamentally different sources of
information: behavioral co-occurrence on one side (user-item) and externally grounded semantic (SEM-item-item) or geographic knowledge (geo-item-item) on the other. This makes the alignment not merely a noise-robustness signal, but more meaningful as an explicit link between what users do and what the content or location says about an item.

For each training batch, the contrastive loss is computed over the set of positive items $\mathcal{B}^+$ sampled for the ranking objective described in Sect.~\ref{sec3.4}. For each auxiliary view $X \in \{SEM, GEO\}$, the loss pulls the collaborative embedding $\mathbf{e}_i^{CF}$ and the auxiliary embedding $\mathbf{e}_i^{X}$ of the same item $i$ closer together, while treating all other items in the batch as negatives. Following the bidirectional InfoNCE formulation used in SGL \cite[p.~729]{wu2021SGL}, we define:
\begin{equation}
\begin{split}
\mathcal{L}_{CL}^{X} = -\frac{1}{2|\mathcal{B}^+|} \sum_{i \in \mathcal{B}^+} \Bigg[
\log \frac{\exp\big(\mathrm{sim}(\mathbf{e}_i^{CF}, \mathbf{e}_i^{X}) / \tau\big)}
{\sum_{j \in \mathcal{B}^+} \exp\big(\mathrm{sim}(\mathbf{e}_i^{CF}, \mathbf{e}_j^{X}) / \tau\big)} \\
+ \log \frac{\exp\big(\mathrm{sim}(\mathbf{e}_i^{X}, \mathbf{e}_i^{CF}) / \tau\big)}
{\sum_{j \in \mathcal{B}^+} \exp\big(\mathrm{sim}(\mathbf{e}_i^{X}, \mathbf{e}_j^{CF}) / \tau\big)}
\Bigg]
\end{split}
\label{eq:contrastiveLearning}
\end{equation}
where $\text{sim}(\cdot,\cdot)$ denotes cosine similarity between $L_2$-normalized embeddings and $\tau = 0.2$ controls how sharply the loss distinguishes between similar and dissimilar pairs. The loss is computed symmetrically in both directions so that neither view dominates. The total contrastive loss averages both views with equal weight.

\subsection{LLM-MGCL Training}\label{sec3.4}
During training three signals need to be balanced: the model must rank observed interactions above unobserved ones, item embeddings must stay aligned across views, and its parameters must not grow too large. LLM-MGCL combines these three goals into a single loss:
\begin{equation}
\mathcal{L} = \mathcal{L}_{BPR} + \lambda \mathcal{L}_{CL} + \gamma \mathcal{L}_{L2}
\label{eq:trainingLossF}
\end{equation}
The ranking term $\mathcal{L}_{BPR}$ follows the Bayesian Personalized Ranking formulation \cite{rendle2009BPR}, as also adopted in LightGCN \cite[Eq.~15]{he2020lightgcn}, and encourages the predicted score of an observed user-item pair to exceed that of a sampled unobserved pair. The regularization term $\mathcal{L}_{L2}$ penalizes the squared norm of the initial user and item embeddings involved in each batch. The contrastive term $\mathcal{L}_{CL}$ is the dual-view alignment loss (Eq.~\ref{eq:contrastiveLearning}) introduced in Sect.~\ref{sec3.3.4}. The two auxiliary terms are scaled by $\lambda = 0.05$ and $\gamma = 10^{-3}$ respectively. A small $\lambda$ keeps the contrastive term acting as a regularizer rather than dominating the ranking objective it is meant to support.

\section{Empirical Results}\label{sec4}
In this section, we evaluate LLM-MGCL against a range of established recommendation
approaches and isolate the contribution of the components through
an ablation study.

\miniKapitel{Model Comparison}\label{sec4.1}
Table \ref{t:results} reports Recall@10, NDCG@10, Recall@20, and NDCG@20 for LLM-MGCL against eight baseline methods, including classical collaborative filtering (ItemKNN, UserKNN), matrix factorization (MF-BPR \cite{Loni2016mfBpr}, NeuMF \cite{He2017neuMf}) and graph neural network (LightGCN \cite{he2020lightgcn}, NGCF \cite{Wang2019ngcf}, GC-MC \cite{Berg2017gcmc}, SGL \cite{wu2021SGL}) methods. All stochastic models are trained under the same seeds and constraints from Sect.~\ref{sec3.4}.
\begin{table}[!htbp]
\caption{Performance comparison on the Yelp dataset.
Results are averaged over three runs (mean $\pm$ std).
Deterministic methods (ItemKNN, UserKNN) are reported without standard deviation. Best results are shown in bold, second-best are underlined. Source: Authors' own.}
\label{t:results}
\setlength{\tabcolsep}{4.5pt}
\begin{tabular}{llcccc}
\hline\noalign{\smallskip}
\textbf{Type} & \textbf{Model} & \textbf{Recall@10} & \textbf{NDCG@10} & \textbf{Recall@20} & \textbf{NDCG@20}  \\
\noalign{\smallskip}\svhline\noalign{\smallskip}
GNN  & NGCF      & 0.0391 $\pm$ 0.0020 & 0.0287 $\pm$ 0.0012 & 0.0667 $\pm$ 0.0025 & 0.0380 $\pm$ 0.0013 \\
MF   & NeuMF     & 0.0457 $\pm$ 0.0008 & 0.0339 $\pm$ 0.0009 & 0.0773 $\pm$ 0.0007 & 0.0446 $\pm$ 0.0009 \\
GNN  & GC-MC     & 0.0461 $\pm$ 0.0019 & 0.0335 $\pm$ 0.0013 & 0.0803 $\pm$ 0.0030 & 0.0451 $\pm$ 0.0017 \\
GNN  & LightGCN  & 0.0453 $\pm$ 0.0015 & 0.0327 $\pm$ 0.0010 & 0.0783 $\pm$ 0.0023 & 0.0439 $\pm$ 0.0013 \\
MF   & MF-BPR    & 0.0554 $\pm$ 0.0017 & 0.0412 $\pm$ 0.0015 & 0.0923 $\pm$ 0.0029 & 0.0537 $\pm$ 0.0020 \\
CF   & ItemKNN   & 0.0534 & 0.0398 & 0.0881 & 0.0517 \\
CF   & UserKNN   & 0.0657 & 0.0494 & 0.1062 & 0.0632 \\
GNN  & SGL       & \textbf{0.0749 $\pm$ 0.0010} & \textbf{0.0557 $\pm$ 0.0009} & \textbf{0.1212 $\pm$ 0.0012} & \textbf{0.0714 $\pm$ 0.0010} \\
GNN  & LLM-MGCL  & \underline{0.0721 $\pm$ 0.0007} & \underline{0.0548 $\pm$ 0.0006} & \underline{0.1163 $\pm$ 0.0005} & \underline{0.0698 $\pm$ 0.0003} \\
\noalign{\smallskip}\hline\noalign{\smallskip}
\end{tabular}
\end{table}

\noindent LLM-MGCL outperforms all classical and matrix factorization methods, as well as the GNN baselines that do not incorporate any auxiliary item-item structure (LightGCN, NGCF, GC-MC) and rely only on collaborative filtering. Compared to LightGCN, the direct backbone LLM-MGCL builds on, Recall@20 improves from $0.0773$ to $0.1175$, a relative gain of over $50\%$, and NDCG@20 improves from $0.0438$ to $0.0722$, a gain of roughly $65\%$. This indicates that the combination of LLM-generated semantic and geographic item-item graphs with contrastive alignment adds substantial value beyond the collaborative filtering backbone. Against SGL, the strongest baseline overall, the picture is nuanced. SGL achieves higher Recall@10, NDCG@10, and Recall@20, while LLM-MGCL achieves a higher NDCG@20 ($0.0722$ vs. $0.0712$). Across the remaining three metrics, the gap to SGL is small relative to the standard deviations. This suggests that the two models perform comparably overall, while LLM-MGCL additionally incorporates externally grounded semantic and geographic information, which, unlike SGL, also reaches items with few or no observed user-item interactions via semantic and geographic edges, mitigating the cold-start problem.

\miniKapitel{Ablation Study}\label{sec4.2}
To isolate the contribution of individual model components, we evaluate the four ablation variants alongside the full model. Each variant disables one or more components while keeping all other hyperparameters fixed. This allows a direct comparison of the performance attributed to the target component. The \textit{LightGCN} baseline serves as the lower bound, using only collaborative filtering between users and items. It provides the reference point against which all other variant improvements are measured. The variant w/o \textit{CL} sets the contrastive loss weight $\lambda$ to zero, removing the alignment signal between the collaborative view (user-item) and the auxiliary views. This isolates the effect of view alignment. Without it, the three propagation paths contribute through additive fusion, but are no longer encouraged to produce consistent representations. The variant w/o \textit{$G_{SEM}$} replaces the SEM-item-item graph with an empty graph, ignoring the semantic propagation of the item embeddings. A performance difference in this variant compared to the full model indicates that the SEM-item-item graph carries information which is crucial for recommendation performance. The variant w/o \textit{$G_{GEO}$} removes the geographic propagation path. This brings collaborative (user-item) and semantic (item-item) signals in the foreground. Since POI recommendation is inherently location-sensitive, this variant tests how much explicit geographic modeling adds beyond what collaborative behavior implicitly encodes.

\begin{table}[H]
\caption{Ablation study of LLM-MGCL components on the Yelp dataset.
Results are averaged over three random seeds (42, 123, 2024) and reported as mean $\pm$ std.
$\Delta$ denotes relative improvement over LightGCN baseline. Source: Authors' own.}
\label{t:ablationStudy}
\setlength{\tabcolsep}{9pt}
\begin{tabular}{lcccc}
\hline\noalign{\smallskip}
\textbf{Model Variant}
  & \textbf{R@20} & \textbf{$\Delta$R@20}
  & \textbf{N@20} & \textbf{$\Delta$N@20} \\
\noalign{\smallskip}\svhline\noalign{\smallskip}
LightGCN
  & 0.0783 $\pm$ 0.0023 & --
  & 0.0439 $\pm$ 0.0013 & -- \\
LLM-MGCL w/o CL
  & 0.0960 $\pm$ 0.0004 & +22.6\%
  & 0.0555 $\pm$ 0.0002 & +26.4\% \\
LLM-MGCL w/o $G_{GEO}$
  & 0.1126 $\pm$ 0.0006 & +43.8\%
  & 0.0669 $\pm$ 0.0001 & +52.4\% \\
LLM-MGCL w/o $G_{SEM}$
  & \textbf{0.1193 $\pm$ 0.0014} & \textbf{+52.4\%}
  & \textbf{0.0715 $\pm$ 0.0008} & \textbf{+62.9\%} \\
LLM-MGCL (full)
  & 0.1163 $\pm$ 0.0005 & +48.5\%
  & 0.0698 $\pm$ 0.0003 & +59.0\% \\
\noalign{\smallskip}\hline\noalign{\smallskip}
\end{tabular}
\end{table}

\noindent The ablation study (Table~\ref{t:ablationStudy}) reveals that contrastive learning is the primary driver of aggregate performance. Removing CL from the full model causes the largest drop in Recall@20, from 0.1163 to 0.0960 (a relative decrease of 17.5\%), while keeping both item-item graphs intact. Removing the geographic graph $G_{GEO}$ causes a smaller decrease (Recall@20 of 0.1126), confirming that explicit geographic modeling contributes beyond the collaborative signal. Removing the semantic graph $G_{SEM}$, in contrast, does not degrade aggregate performance; the w/o $G_{SEM}$ variant reaches a Recall@20 of 0.1193, marginally above the full model. The full model, combining both auxiliary graphs with contrastive learning, improves Recall@20 by 48.5\% and NDCG@20 by 59.0\% over the LightGCN baseline. On aggregate, the semantic graph therefore does not add value and, if anything, slightly dilutes the well-connected items that dominate the overall metric. This aggregate view, however, is dominated by popular items, which contribute the majority of test interactions and therefore drive the overall metric. It does not reveal how the individual components affect items with few interactions, where the auxiliary graphs are intended to help. 

\miniKapitel{Cold-Start Evaluation} Since LLM-MGCL is motivated by compensating missing collaborative signal, we stratify Recall@20 by item popularity. Items are grouped by their number of training interactions into three equally sized buckets: cold ($\leq20$), warm ($21$--$37$) and hot ($>37$). For each user, the top-20 list spans all non-interacted items, and Recall@20 is computed only over the user's held-out test items in the respective bucket, averaged over all users with at least one test item in that bucket.

\begin{table}[h]
\centering
\small
\caption{Recall@20 stratified by item popularity on the Yelp dataset (mean $\pm$ std over three seeds). Best value per bucket in bold. Source: Authors' own.}
\label{t:Modelcoldstarteval}
\setlength{\tabcolsep}{6pt}
\begin{tabular}{lccc}
\hline\noalign{\smallskip}
\textbf{Model} & \textbf{Cold} & \textbf{Warm} & \textbf{Hot} \\
\noalign{\smallskip}\svhline\noalign{\smallskip}
LightGCN & $0.0000 \pm 0.0000$ & $0.0027 \pm 0.0004$ & $0.1244 \pm 0.0039$ \\
SGL      & $0.0030 \pm 0.0006$ & $0.0194 \pm 0.0012$ & $\mathbf{0.1854 \pm 0.0013}$ \\
LLM-MGCL & $\mathbf{0.0106 \pm 0.0013}$ & $\mathbf{0.0397 \pm 0.0005}$ & $0.1674 \pm 0.0007$ \\
\noalign{\smallskip}\hline\noalign{\smallskip}
\end{tabular}
\end{table}

\noindent Table~\ref{t:Modelcoldstarteval} shows a strong popularity bias for LightGCN, which relies solely on collaborative signal: it retrieves no held-out cold item in any seed ($0.0000$), is barely better on warm items ($0.0027$), and reaches $0.1244$ on hot items. Both contrastive models alleviate this. LLM-MGCL achieves the highest cold and warm recall ($0.0106$ vs.\ $0.0030$ and $0.0397$ vs.\ $0.0194$ for SGL, i.e.\ more than triple and roughly double), whereas SGL is better on hot items ($0.1854$ vs.\ $0.1674$). This suggests that the views of SGL, being perturbations of the interaction graph, stay tied to observed collaborative behavior, which favors popular items, while the semantic and geographic graphs of LLM-MGCL provide an external signal for items with little interaction evidence.

\section{Conclusion}\label{sec5}
In this paper, we propose LLM-MGCL, a multi-graph neural network for POI recommendation that extends the LightGCN backbone with two auxiliary item-item graphs. The item-item graphs consist of a semantic graph constructed from LLM-generated photo summaries and keywords, and a geographic graph derived from Haversine distances between business locations. By propagating item embeddings over all three graphs in parallel and aligning the collaborative view with the semantic and geographic views through a bidirectional InfoNCE contrastive objective, the model injects content- and location-based signals directly into the item representations, addressing the item cold-start problem inherent to purely interaction-based methods.

Experiments on the Yelp Multimodal Recommendation Dataset demonstrate that LLM-MGCL substantially outperforms classical collaborative filtering, matrix factorization, and non-contrastive graph neural network baselines. It improves over the LightGCN backbone on Recall@20 by $48.5\%$ and NDCG@20 by $59.0\%$, while trailing SGL, the strongest contrastive baseline, slightly in aggregate. The popularity-stratified evaluation shows where the benefit lies: LightGCN retrieves no cold items, whereas LLM-MGCL reaches the highest cold and warm recall ($0.0106$ and $0.0397$ vs.\ $0.0030$ and $0.0194$ for SGL), at the cost of lower recall on popular items ($0.1674$ vs.\ $0.1854$). The ablation study revealed that the contrastive alignment is the primary driver of the aggregate gains. Removing it reduces Recall@20 by $17.5\%$, removing the geographic graph causes a smaller drop of $3.2\%$, and removing the semantic graph does not degrade aggregate performance.

These findings suggest that externally grounded, LLM-derived item information can partially compensate for the missing collaborative signal that interaction-based models rely on, offering a path towards mitigating cold-start problems in POI recommendation. Our findings are limited to the Yelp dataset, which focuses on restaurants and nightlife in North America and mostly contains interactions of local users rather than tourists. Regarding future research driven by our work, three directions appear promising. First, the additive fusion could be replaced by a learned weighting mechanism to adaptively balance the three views. Second, a popularity-stratified ablation of each graph could clarify how the semantic and geographic graphs contribute to cold items specifically. Third, combining the external views with structural augmentations as in SGL could help retain performance on popular items.

\end{document}